# Two-band BCS superconductivity in Ba(Fe$_{0.9}$Co$_{0.1}$)$_2$As$_2$

E.G.Maksimov[1], A.E.Karakozov[2], B.P.Gorshunov[3,4], V.S.Nozdrin[3], A.A.Voronkov[3], E.S.Zhukova[3,4], S.S.Zhukov[5], Dan Wu[6], M. Dressel[6], S. Haindl[7], K. Iida[7], B. Holzapfel[7]

[1] P.N.Lebedev Physical Institute, Russian Academy of Sciences, 119991 Moscow, Russia

[2] Institute of High Pressure Physics, Russian Academy of Sciences, Troitsk, Moscow region, 142190 Russia

[3] A.M.Prokhorov Institute of General Physics, Russian Academy of Sciences, 119991 Moscow, Russia

[4] Moscow Institute of Physics and Technology (State University), Institutskii per. 9, 141700, Dolgoprudny, Moscow Region, Russia

[5] Volgograd State Technical University, Lenin avenue 28, 400131 Volgograd, Russia

[6] 1. Physikalisches Institut, Universität Stuttgart, Pfaffenwaldring 57, 70550 Stuttgart, Germany

[7] IFW Dresden, Institute for Metallic Materials, P. O. Box 270116, 01171 Dresden, Germany

**Abstract**

The conductivity and permittivity optical spectra of iron-pnictide Ba(Fe$_{0.9}$Co$_{0.1}$)$_2$As$_2$ film (T$_c$=20 K) are analyzed. In the superconducting state, at all temperatures up to T$_c$ the temperature dependences of the magnetic field penetration depth and of the superconducting condensate density are well described within the generalized two-band BCS model with intraband and interband pairing interactions considered. It is shown that the smaller superconducting energy gap 2$\Delta$ = 3.7 meV develops in the electronic subsystem while the larger gap 2$\Delta \geq$ 7 meV opens in the hole subsystem. The normal state parameters (plasma frequencies and scattering rates) of electron and hole conduction bands are determined. At all temperatures the obtained data are consistent with the results of electronic photoemission experiments on Ba(Fe$_{1-x}$Co$_x$)$_2$As$_2$.

**Introduction**

These days situation with the recently discovered [1] iron-based superconductors looks similar to that with cuprates in the late eighties. To get insight into the mechanism of superconductivity a lot of efforts are made to obtain reliable information on the most fundamental properties of iron pnictides like the superconducting (SC) energy gaps in different bands, the kind of excitations that assist Cooper-pair formation, the temperature dependences of the magnetic field penetration depth $\delta(T)$ and of the SC condensate, etc. Two years since the discovery, the experimental data and theoretical approaches are rather ambiguous and contradictory. There exists broad scattering in the obtained values of the ratio 2$\Delta$/T$_c$=2-10 and of other quantities that characterize iron-based compounds [2,3,4]. No consensus exists on the pairing symmetry. One of the most widely discussed models of superconductivity is the s$^\pm$ model [5]. It considers electronic and hole bands at the M and $\Gamma$ points of the Brillouin zone, respectively, that possess s-type SC order parameters with phases of opposite signs. Experimentally, however, both kinds of energy gaps are reported, s-type [4,6] as well as d-type [3,7,8]. As for the coupling mechanism, in spite of a sizable isotope effect [9,10], substantial part of researchers believe that Cooper pairs are formed not due to phonons but due to magnetic excitations.

Unlike single-band BCS and cuprate superconductors, or two-band $MgB_2$, Fe-based compounds are multi-band systems. In the $BaFe_2As_2$-based compounds, in addition to at least two electron [near the point X $(0,\pm\pi)$] and two hole (at the Γ-point) bands [11], one more hole band can exist close to the X-point whose presence depends crucially on the structural details [12,13]. The multi-band nature of iron pnictides is one of the factors that complicate the analysis of their physical properties. Effective means for studying charge-carriers dynamics in different bands in the normal and in the SC states are provided by optical spectroscopy. Up to now quite a number of optical investigations have been carried out, based mainly on measurements of the reflectivity spectra $R(\nu)$ in the infrared range utilizing Fourier-transform spectrometers. Although the SC transition is clearly displayed in the spectra $R(\nu)$, the extracted quantitative characteristics (like the values of SC gaps, their symmetries, quantities and assignment to certain bands) and corresponding conclusions are rather controversial. The reason lies partly in the relatively low energy (frequency) scale set by critical temperatures in iron pnictides, $T_c \leq 55$ K, $\nu \leq k_B T_c/hc \leq 150$ cm$^{-1}$ ($k_B$ – Boltzmann constant, h – Planck's constant, c – speed of light). Here Fourier-transform spectrometers loose their efficiency, especially when measuring superconductors whose reflectivity is very close to 100% [$R \geq 99.8\%$ for $Ba(Fe_{0.9}Co_{0.1})_2As_2$ at $\nu=5$ cm$^{-1}$, see below]. At the same time the frequencies from 50 cm$^{-1}$ down to 1 cm$^{-1}$ are effectively covered by continuous-wave terahertz (THz) spectrometers based on BWOs as coherent radiation sources [14] (BWO – backward-wave oscillator). These spectrometers allow for direct (no Kramers-Kronig analysis needed) quantitative measurements of the conductivity $\sigma(\nu)$ and permittivity $\varepsilon'(\nu)$ spectra of conductors and superconductors [15].

## Experimental Data

By using a BWO spectrometer we have performed first direct measurements of the THz (4 cm$^{-1}$ – 45 cm$^{-1}$) spectra of $\sigma(\nu)$ and $\varepsilon'(\nu)$ of a $Ba(Fe_{0.9}Co_{0.1})_2As_2$ film ($T_c=20$ K) on a dielectric $(La,Sr)(Al,Ta)O_3$ substrate. The THz measurements were extended up to 35000 cm$^{-1}$ by employing Fourier-transform spectrometer and optical ellipsometer. The main experimental findings are reported in [16]: (i) a SC gap $2\Delta_0=(3.7\pm0.3)$ meV and a large intra-gap absorption were directly observed and (ii) the temperature dependence of the magnetic field penetration depth $\delta(T)$ with a low-temperature value $\delta(T \to 0) = (3600 \pm 500)$ Å was determined. In this report, we present an advanced and more in-depth analysis of these data. We find that the normal state electrodynamics of $Ba(Fe_{0.9}Co_{0.1})_2As_2$ is determined by electron and hole bands with plasma frequencies and scattering rates $\nu_{pl}^{el} = 8570$ cm$^{-1}$, $\gamma^{el} = 110$ cm$^{-1}$ and $\nu_{pl}^{h} = 5850$ cm$^{-1}$, $\gamma^h = 400$ cm$^{-1}$ (at T = 30 K), respectively. In addition to the two interband transitions at ≈4000 cm$^{-1}$ and ≈21000 cm$^{-1}$, we observe a resonance absorption at ≈1000 cm$^{-1}$ of yet unknown origin. In the SC-state, the smaller SC gap with $2\Delta = 3.7$ meV develops in the electronic band while the larger gap of $2\Delta \geq 7$ meV opens in the hole band. In the whole temperature interval $0 < T < T_c$ the temperature behavior of the magnetic field penetration depth and of the SC condensate density can be well reproduced within a moderately strong coupling BCS two-band model involving electronic and hole bands with intraband (coupling constant $\lambda \approx 0.45$ in one of the bands) and interband ($\lambda \approx 0.1$) pairing interactions.

## Analysis
### Normal state

Fig.1(a) shows raw data from [16] – the spectra of reflection coefficient of $Ba(Fe_{0.9}Co_{0.1})_2As_2$ film on a $(La,Sr)(Al,Ta)O_3$ substrate measured in the normal (30 K) and in the SC (5 K) states. Besides relatively narrow features between 100 cm$^{-1}$ and 1000 cm$^{-1}$ due to phonons in the substrate, there are two broad bumps above 1000 cm$^{-1}$ coming from interband transitions in Ba(Fe$_{0.9}$Co$_{0.1}$)$_2$As$_2$ that can be modeled with Lorentzians:

$$\sigma^*(\nu) = \frac{0.5 f \nu}{\nu \gamma_L + i(\nu_0^2 - \nu^2)} \qquad (1),$$

with $f = \Delta\varepsilon \nu_0^2$ the oscillator strength, $\Delta\varepsilon$ – the dielectric contribution, $\nu_0$ – the eigenfrequency and $\gamma_L$ – the damping. An upturn of the reflectivity below 1000 cm$^{-1}$ is caused by itinerant charge carriers - electrons and holes in various bands of Ba(Fe$_{0.9}$Co$_{0.1}$)$_2$As$_2$. Photoelectronic emission experiments described in [17] provide important information on the relative spectral weights (plasma frequencies $\nu_{pl}$) of the two electronic (at the M point) and two hole (holes A and B, according to [17], at the Γ-point) conduction bands. Assuming the 2d nature of all bands the carriers concentration that is proportional to the volume within the Fermi surface can be considered as proportional to $k_F^2$ leading to $k_F \sim \nu_{pl}$. Taking into account the main contribution of the hole B-component, for our case x(Co)=0.1 we get $k_F^{el}/k_F^{h} = \nu_{pl}^{el}/\nu_{pl}^{h} \approx 1.5$, clearly demonstrating that in the normal state the electronic contribution to the optical and DC conductivity of Ba(Fe$_{0.9}$Co$_{0.1}$)$_2$As$_2$ dominates over the hole contribution, in accordance with the transport and optical data of [18,19].

We model the response of free carriers with the standard Drude expression for complex conductivity:

$$\sigma^*(\nu) = \sigma_1(\nu) + i\sigma_2(\nu) = \frac{\sigma_0 \gamma^2}{\gamma^2 + \nu^2} + i \frac{\sigma_0 \nu \gamma}{\gamma^2 + \nu^2} \qquad (2),$$

where $\sigma_1$ is the real part and $\sigma_2 = \nu(\varepsilon_\infty - \varepsilon')/2$ the imaginary part of the conductivity, $\varepsilon_\infty$ is the high-frequency dielectric constant, $\sigma_0 = \frac{\nu_{pl}^2}{2\gamma}$ is the DC conductivity [20]. Using photoemission data [17] for relative contributions of all four (two electronic and two hole) bands and employing expressions (1,2) we fit the normal state reflectivity spectrum of the Ba(Fe$_{0.9}$Co$_{0.1}$)$_2$As$_2$ film on a substrate (grey line in Fig.1a). Fig.1(b) shows the resultant conductivity spectrum of the film (thick line) together with the separate components (thin lines). The unambiguously determined parameters of all terms are listed in Table I. It is important that while processing the reflectivity spectrum we had to introduce a resonance absorption at ≈1000 cm$^{-1}$ (Lorentzian in Table I). A similar absorption feature was detected in [21,22], and also in undoped BaFe$_2$As$_2$ [19,23]. According to a theoretical analysis [24,25], its origin could be connected with low-energy interband transitions in electronic and hole sub-bands.

Fig.1(b) and Table I again show that the main contribution to the low-frequency and DC conductivity of Ba(Fe$_{0.9}$Co$_{0.1}$)$_2$As$_2$ at T=30 K comes from the electronic component. We have obtained the temperature dependence of its parameters by processing higher-temperature reflectivity spectra. The results are shown in Fig.2. Clearly, the increase of the DC resistivity (inset in Fig.1b) from ρ(30 K)≈0.1 Ωcm to ρ(300 K)≈0.2 Ωcm is mainly caused by an increase of the electronic scattering rate (Fig.2a). Considering bosons (for example, phonons) as scatterers, we can use expression $\gamma = 2\pi\lambda T$ [26] to estimate the coupling constant λ: with γ(300 K) = (490±50) cm$^{-1}$ we obtain λ ≈ 0.45. It is well known [26] that the temperature variation of the charge carriers scattering rate should necessarily be linked to its frequency variation. In our analysis, however, we did not detect any dispersion of the electronic γ. The reason lies in its significantly different sensitivity to changes of the temperature and of the frequency [26]. Indeed, for T=0 the maximal value of the scattering rate at high frequencies ($\nu > 4\langle\nu\rangle$) is $\gamma_{max}$ = $\lambda\langle\nu\rangle$, where $\langle\nu\rangle$ is the mean frequency of bosons, which for the case of phonons in

Ba(Fe$_{0.9}$Co$_{0.1}$)$_2$As$_2$ amounts in $\langle v \rangle/c \approx 180$ cm$^{-1}$ [27,28]. With $\lambda \approx 0.45$, we find that the complete drop of electronic $\gamma$ between the lowest and the highest frequencies is not more than $\approx 80$ cm$^{-1}$. (At higher temperatures this drop is even smaller [26]). To experimentally detect this relatively small change of $\gamma$ in a broad frequency interval (from $v \approx 10$ cm$^{-1}$ to $v \geq 10^3$ cm$^{-1}$) is rather difficult, also in conditions when the spectra of Ba(Fe$_{0.9}$Co$_{0.1}$)$_2$As$_2$ include contributions from other bands and from low-energy interband transitions.

## Superconducting state

Fig.3 shows the $\sigma(v)$ and $\varepsilon'(v)$ spectra of Ba(Fe$_{0.9}$Co$_{0.1}$)$_2$As$_2$ in the SC-state, at T = 5 K [16]. The decrease of $\sigma(v)$ down to zero at $v = 30$ cm$^{-1}$ indicates a SC energy gap $2\Delta/hc = 30$ cm$^{-1}$. It is clear that this gap opens in the electron band since the value $\sigma \approx 5 \times 10^3$ $\Omega^{-1}$cm$^{-1}$ right above the gap frequency is too high to be associated with the holes (see also Table I). We were not able to find the value of SC energy gap in the hole condensate by analyzing the reflectivity spectra because of relatively small contribution of holes to the conductivity. On the other hand, this value can be determined from the temperature behavior of the magnetic field penetration depth at T $\geq$ T$_c$/2, see below.

We note a puzzling result reported in [16] - a large absorption at the lowest frequencies, $v < 10$ cm$^{-1}$, which is observed also in other iron pnictides ([2] and references therein). There are attempts to associate this absorption with a pair breaking effects within the s$^\pm$ model (see for example [29]). However, our results map out the standard s$^\pm$ model that takes into account only the interband pairing interaction. Within this model the following relation between the gap values $\Delta_i$ and densities of states (DOS) N$_i$ in different bands should hold: $\frac{\Delta_1}{\Delta_2} = \sqrt{\frac{N_2}{N_1}}$ [30]. According to the band structure calculations [5] and specific heat experiments [31] the electronic DOS in Ba(Fe$_{0.9}$Co$_{0.1}$)$_2$As$_2$ is by almost two times smaller than the total DOS in the hole bands. This implies that the SC-gap in the electronic condensate should be larger than the gap in the hole subsystem, at variance with our results. Arguments against a simple s$^\pm$ model were given also in [32].

We tried to describe the $\sigma(v)$ and $\varepsilon'(v)$ spectra of Ba(Fe$_{0.9}$Co$_{0.1}$)$_2$As$_2$ at the lowest temperature T = 5 K within the BCS model involving arbitrary impurity scattering [33]. We took the band parameters listed in Table I. We know the SC gap value $2\Delta_{el}/hc=2\Delta_1/hc=30$ cm$^{-1}$ (Fig.3) for electrons and for the holes we find (see below) the value $2\Delta_h/hc=2\Delta_2/hc=60$ cm$^{-1}$ that agrees with specific heat experiments [31]. Combining contributions of electrons and holes additively, we obtain the result presented in Fig.3. It is clearly seen that the BCS calculations reproduce the low-temperature spectra well (except the large below-gap absorption mentioned above which could also be a reason of a slight difference between calculated and measured values of $\varepsilon'$). This inspired us to further employ the BCS formalism to describe the temperature dependences of the magnetic field penetration depth $\delta(T)$ and of the SC condensate density $\rho_s(T) \sim 1/\delta^2(T)$ determined in [16] on the basis of directly measured dielectric permittivity of Ba(Fe$_{0.9}$Co$_{0.1}$)$_2$As$_2$. For a BCS superconductor with impurities the penetration depth is expressed as [34]:

$$\frac{1}{\delta_i^2} = (2\pi)^3 \frac{\sigma_{0,i}}{c^2}\left(\frac{\Delta_i}{\hbar}\right)\left\{th\frac{\Delta_i}{2T} - \frac{2}{\pi}\int_0^\infty \frac{(\gamma_{imp}^i/\Delta_i)dx}{\sqrt{x^2+1}} \frac{th\left(\frac{\Delta_i}{2T}\sqrt{x^2+1}\right)}{x^2+(\gamma_{imp}^i/\Delta_i)^2}\right\} \qquad (3).$$

Here i = 1,2 stands for electrons and holes. Using Eq. (3) with the parameters of electron and hole bands summarized in Table I and SC gaps $2\Delta_{el}/hc=30$ cm$^{-1}$ and $2\Delta_h/hc=60$ cm$^{-1}$ [31] for T$\rightarrow$0 we get $\rho_{s,h} \approx 0.2\rho_{s,e}$. This means that at T = 5 K the contribution to the SC condensate from the hole bands are noticeably smaller than the contribution from the electronic bands. For T → 0 we also obtain $\delta_{el}(0) \approx 3700$ Å, $\delta_h(0) = 7530$ Å and $\delta_{tot}(0) = \left[1/\delta_{el}^2(0) + 1/\delta_h^2(0)\right]^{-1/2} \approx 3300$ Å, the value $\delta_{tot}(0)$ practically coinciding with the experimentally measured [16] $\delta_{exp}=3600$ Å. Note that assigning a larger gap $2\Delta/hc \geq 60$ cm$^{-1}$ to the electron band and the smaller one $2\Delta/hc=30$ cm$^{-1}$ to the hole subsystem leads to significantly lower value $\delta_{tot}(0) \leq 2400$ Å providing another evidence that the smaller gap in Ba(Fe$_{0.9}$Co$_{0.1}$)$_2$As$_2$ develops in the electron bands.

Our analysis showed that the higher temperature, T>5 K, behavior of $\delta(T)$ and $\rho_s(T)$ could not be reproduced within either the α-approximation of the BCS approach [33] (see dashed lines in Fig.4a,b) or simple s$^{\pm}$ model. In this regard we make one step further in our BCS analysis and introduce an interband pairing interaction. This makes the situation in Ba(Fe$_{1-x}$Co$_x$)$_2$As$_2$ similar to that in MgB$_2$. It is known, that MgB$_2$ is a two-band superconductor with intraband pairing interaction defining the gap values and with a certain interband coupling leading to a single transition temperature of the material. The SC gaps in MgB$_2$ have different values, which depend differently on the doping level: adding Al causes a decrease of the larger gap while the smaller gap stays unchanged ([35] and references therein). This is similar to BaFe$_{1-x}$Co$_x$As$_2$ [2], where the smaller gap $2\Delta/hc = 30$ cm$^{-1}$ is practically independent on x(Co) and the larger gap changes with x(Co) proportional to T$_c$. It is also important that the temperature variation of the smaller gap in MgB$_2$ cannot be reproduced within the α-approximation but needs to consider the interband interaction [35].

Introducing the interband coupling we use the following expression for the temperature dependence of the SC gaps in Ba(Fe$_{0.9}$Co$_{0.1}$)$_2$As$_2$ [35]:

$$\ln \frac{\Delta_1(T)}{\Delta_1(0)} = -2\int_0^\infty dy \frac{f\left\{\frac{\Delta_1(T)}{T}\sqrt{1+y^2}\right\}}{\sqrt{1+y^2}} + \lambda_{12}\left\{\frac{\Delta_2(T)}{\Delta_1(T)} - \frac{\Delta_2(0)}{\Delta_1(0)}\right\} \quad (4).$$

Here $\lambda_{12}$ – is the coupling constant between the electron and the hole bands and f(x) is the Fermi distribution function. With a single fitting parameter $\lambda_{12}$, we determined the proper and unambiguous temperature dependences of SC gaps in electronic and hole bands and the value $\lambda_{12}=0.1$ that perfectly describe the dependences $\delta(T)$ and $\rho_s(T)$, as seen in Fig.4a,b. We would like to stress that the obtained accordance is realized with the value of the SC gap in the hole condensate $2\Delta_h/hc(0) = 60$ cm$^{-1}$ mentioned above. This value can thus be considered as unambiguously determined by the functional dependence of the penetration depth on the temperature at T ≥ 0.6T$_c$ = 12 K where the contribution to the SC condensate of the hole component prevails over the contribution of electronic component. At T ≤ 0.6T$_c$ = 12 K the penetration depth and the SC-condensate density are determined mainly by the electronic component. Different roles that the electronic and hole components play in the formation of the SC phase of Ba(Fe$_{0.9}$Co$_{0.1}$)$_2$As$_2$ above and below 0.6T$_c$ is related with the temperature dependence of the electronic gap which at T>0.6T$_c$ is considerable smaller than the gap given by α-approximation (dashed line in Fig.4c). A characteristic dip in electronic gap around T ≈ 0.7T$_c$ = 14 K (Fig.4c) is similar to the one observed in the tunnel experiments on MgB$_2$ [35]. We note that in the above analysis the α-approximation has been used for the temperature dependence of the gap in the hole band.

Neglecting the second term in the brackets of Eq.(4) leads to a standard BCS solution for the electronic gap $\Delta_1(T)$ with $2\Delta_1(0)/T_c = 3.51$ and with the critical temperature T$_c$ = 11.7 K. It is intriguing that basically the same value T$_c$ = 11.4 K is obtained by the well known expression

$$T_c = \frac{\Omega}{1.2}\exp\left[-\frac{1+\lambda}{\lambda}\right] \quad (5)$$

with the effective boson frequency $\Omega$ of the order of the mean phonon frequency in Ba(Fe$_{1-x}$Co$_x$)$_2$As$_2$, i.e. $\Omega$ = 30 meV/hc ≈ 240 cm$^{-1}$ [27], and with the electron-phonon coupling constant $\lambda$ ≈ 0.45 estimated above.

## Conclusions

In conclusion, the analysis of optical data of Ba(Fe$_{0.9}$Co$_{0.1}$)$_2$As$_2$ shows that in the SC state the smaller energy gap 2$\Delta$ = 3.7 meV develops in the electronic subsystem while the larger gap 2$\Delta$ ≥ 7 meV opens in the hole subsystem. At T = 5 K the spectra of conductivity and permittivity can be well described within the BCS model with relative normal state spectral weights of electron and hole bands in accordance with electronic photoemission data. In the full range 0 < T < T$_c$, the temperature dependences of the magnetic field penetration depth and of the SC condensate density are well reproduced within the generalized two-band BCS model with intraband and interband pairing interactions, but cannot be explained by simple s$^\pm$ model. In the normal state, the electrodynamic properties of Ba(Fe$_{0.9}$Co$_{0.1}$)$_2$As$_2$ are determined by electronic and hole systems with plasma frequencies and scattering rates $v_{pl}^{el}$ = 8570 cm$^{-1}$, $\gamma^{el}$ = 110 cm$^{-1}$ and $v_{pl}^{h}$ = 5850 cm$^{-1}$, $\gamma^{h}$ = 400 cm$^{-1}$ (T = 30 K). The low-frequency behavior and the DC conductivity are determined mainly by the electronic condensate for which the scattering rate decreases from 500 cm$^{-1}$ at room temperature to 110 cm$^{-1}$ at 30 K due to bosonic interaction with the coupling constant $\lambda$ ≈ 0.45. Besides two interband transitions at approximately 4000 cm$^{-1}$ and 21000 cm$^{-1}$, a resonance absorption is detected at around 1000 cm$^{-1}$ of yet unknown origin.

Acknowledgements. We thank A.Kordyuk and D.Evtushinsky for helpful discussions of the photoemission data. D.W. is grateful to receive support by the Alexander von Humboldt-Foundation. The work was supported by the RAS Programs for fundamental research "Strongly correlated electrons in solids and solid structures" and "Problems of Radiophysics" and Russian Foundation for Basic Research (grant № 09-02-0056-a).

Table I. Normal state of Ba(Fe$_{0.9}$Co$_{0.1}$)$_2$As$_2$, T = 30 K: parameters of carriers in electronic and hole bands (DC conductivities $\sigma_{dc}$, plasma frequencies $\nu_{pl}$, scattering rates $\gamma$) and of resonances (eigenfrequencies $\nu_0$, oscillator strengths f, dielectric contributions $\Delta\varepsilon$, dampings $\gamma_L$). Superconducting state, T=5 K: plasma frequencies of superconducting condensates in electronic and hole bands calculated according to BCS model [31] taking into account relative normal state spectral weights of the bands according to photoemission experiments [17].

| | Normal state, T=30 K | | | | | | |
|---|---|---|---|---|---|---|---|
| | Electron band | Hole band B | Hole band A | | Interband transition I | Interband transition II | Lorentzian at ≈1000 cm$^{-1}$ |
| $\sigma_{dc}$ ($\Omega^{-1}$cm$^{-1}$) | 11 000 | 1 260 | ≈220 | $\nu_0$ (cm$^{-1}$) | 4 340 | 21 400 | 1 070 |
| $\nu_{pl}$ (cm$^{-1}$) | 8 570 | 5 500 | ≈2000 | f$^{1/2}$ (cm$^{-1}$) | 28 000 | 100 000 | 9 400 |
| $\gamma$ (cm$^{-1}$) | 110 | 400 | ≈300 | $\gamma_L$ (cm$^{-1}$) | 6 200 | 55 000 | 1 750 |
| | | | | $\Delta\varepsilon$ | 40 | 20 | 80 |
| Superconducting state, T=5 K | | | | | | | |
| $\nu_{pl}^{SC}$ (cm$^{-1}$) | 4300 | 2300 | ≈850 | | | | |

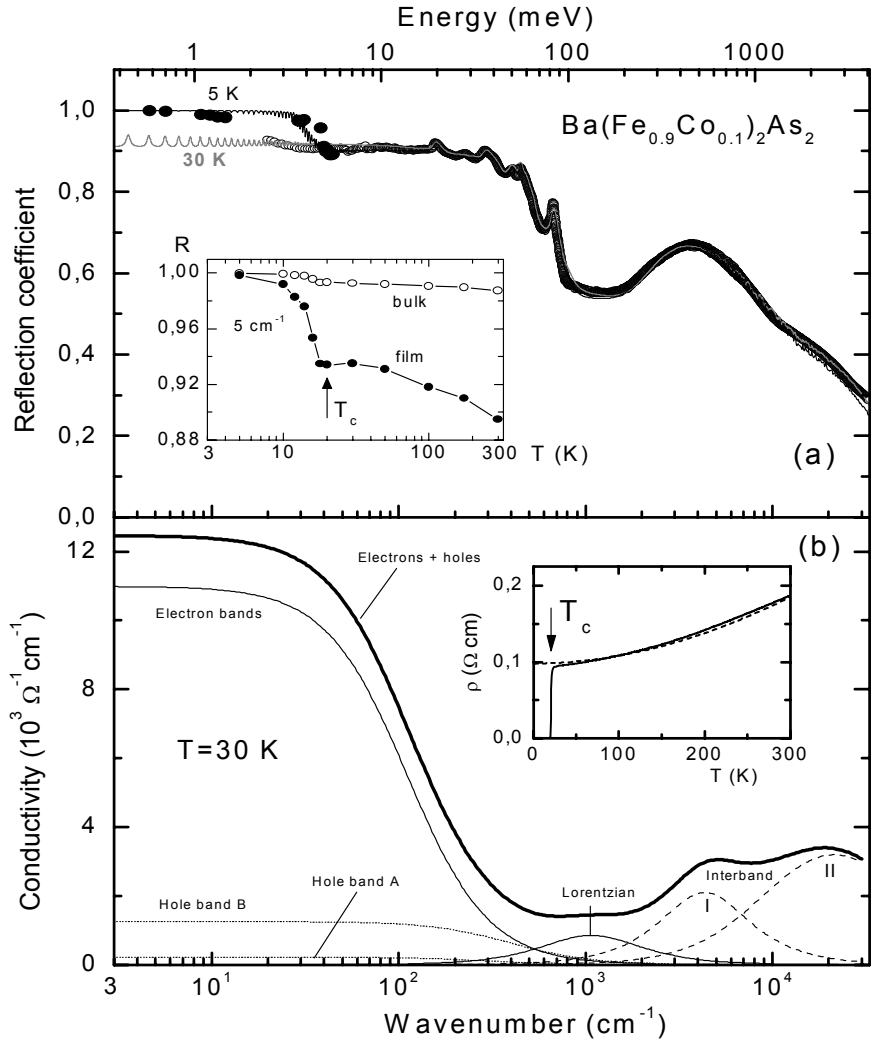

Fig.1. (a). Spectra of reflection coefficient of Ba(Fe$_{0.9}$Co$_{0.1}$)$_2$As$_2$ film on (La,Sr)(Al,Ta)O$_3$ substrate at T=30 K [normal state of Ba(Fe$_{0.9}$Co$_{0.1}$)$_2$As$_2$, open dots] and at T=5 K (superconducting state, black dots) [16]. Inset: temperature dependences of reflection coefficients of the film on a substrate and of the bulk Ba(Fe$_{0.9}$Co$_{0.1}$)$_2$As$_2$ (calculated using determined values of conductivity and permittivity of the film) at a fixed frequency 5 cm$^{-1}$. Grey solid line shows the result of a fit of the reflectivity at 30 K with the obtained conductivity spectrum presented on panel (b) (thick line) together with the contributions of free carriers (electrons, holes) and resonance absorption (thin lines). Black line – reflectivity in the SC state calculated on the basis of conductivity and permittivity shown in Fig.3. Inset on panel (b) shows the temperature dependence of the DC resistivity, the dashed line corresponds to the resistivity of electronic component calculated on the basis of parameters (Fig.2) determined from the optical spectra.

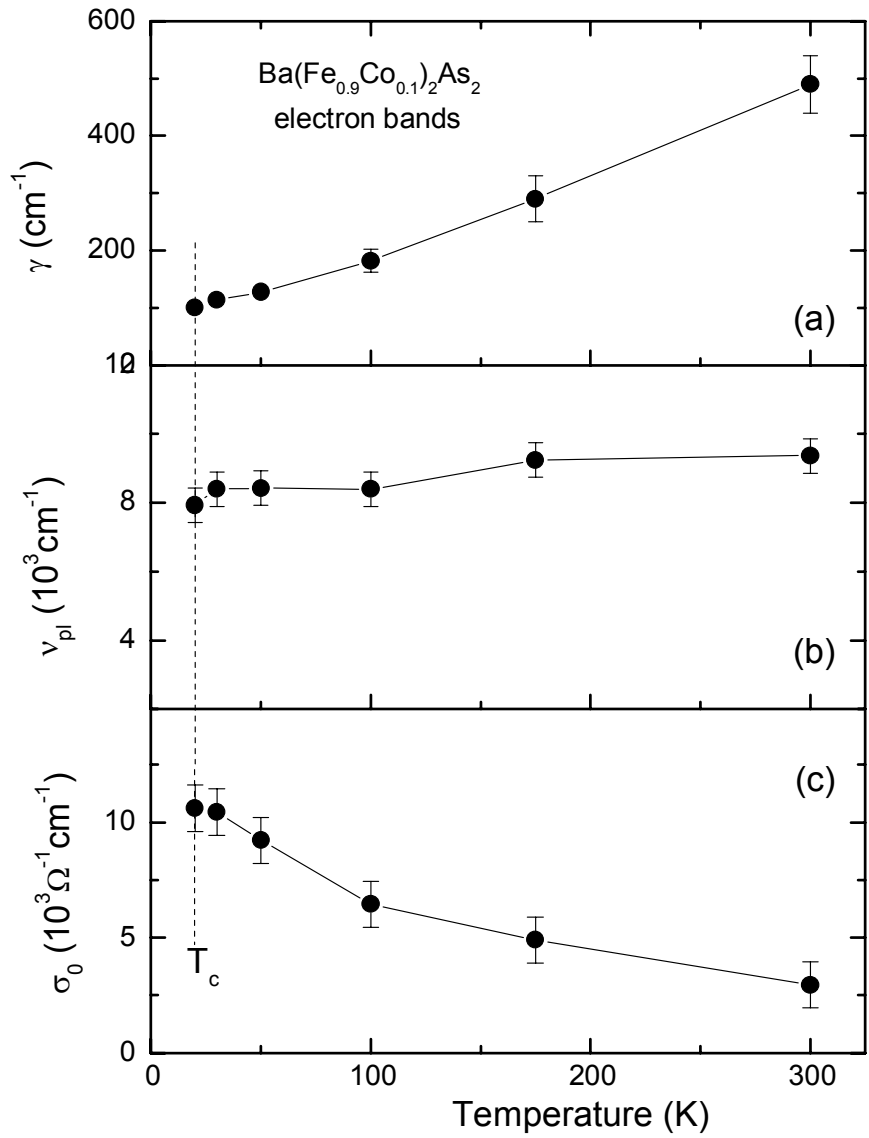

Fig.2. Temperature dependences of the parameters for the electronic free-carrier component in the normal state of Ba(Fe$_{0.9}$Co$_{0.1}$)$_2$As$_2$, determined by fitting the reflectivity spectra shown in Fig.1: (a) scattering rate, (b) plasma frequency and (c) DC conductivity.

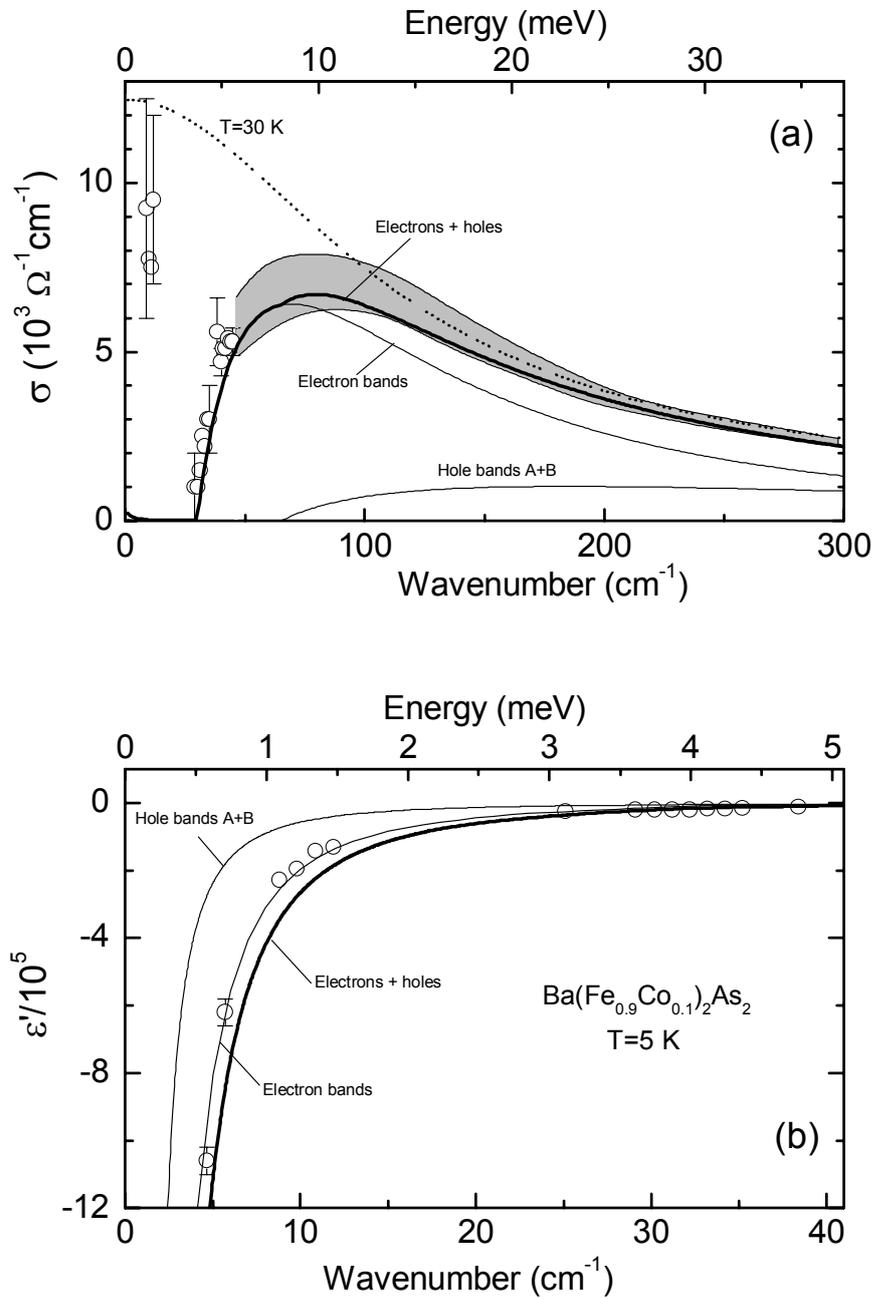

Fig.3. Spectra of (a) conductivity and (b) permittivity of $Ba(Fe_{0.9}Co_{0.1})_2As_2$ in the superconducting state (T=5 K) [16]. Dots – direct measurements on the BWO-spectrometer. Grey area on panel (a) shows conductivity spectrum obtained by fitting the spectrum of reflection coefficient (Fig.1a) taking into account uncertainties of the fitting procedure and directly measured values of THz conductivity and permittivity. Thin lines show contributions from electronic and hole (A and B according to [17]) bands calculated within the BCS model [31], thick lines – total contributions of electrons and holes. Dotted line – conductivity spectrum in the normal state at T=30 K.

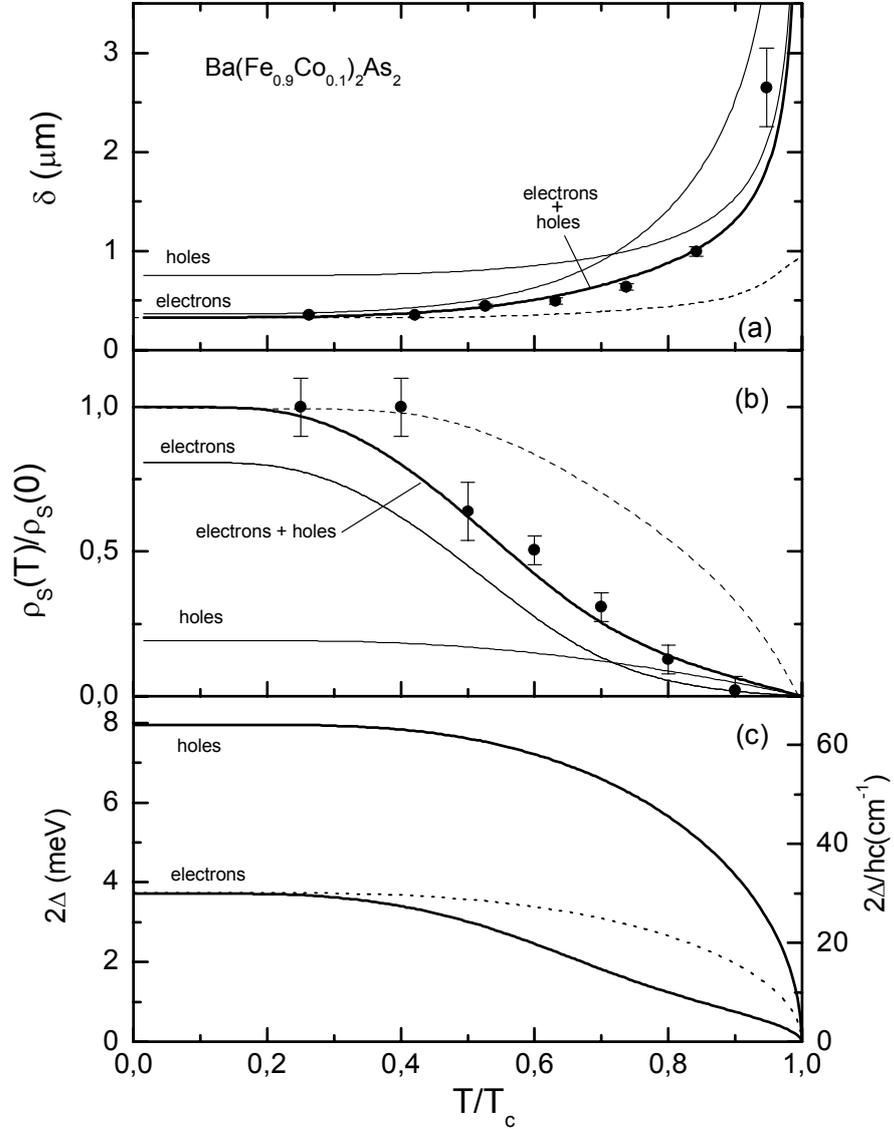

Fig.4. Temperature dependences of magnetic field penetration depth (a), spectral weight of the superconducting condensate (b) and energy gaps in electronic and hole bands of $Ba(Fe_{0.9}Co_{0.1})_2As_2$. Dots – experiment [16]. Lines – calculation according to extended BCS two-band model with electron-boson coupling constants: intraband $\lambda=0.5$, interband $\lambda=0.1$. Dashed lines on panels (a) and (b) show calculations within the α-approximation of the BCS model (additive contributions of electronic and hole bands). Dashed line on panel (c) – temperature dependence of the gap in electronic subsystem without interband coupling.